# The property of $\kappa$-deformed statistics for a relativistic gas in an electromagnetic field: $\kappa$ parameter and $\kappa$-distribution

Guo Lina, Du Jiulin[*] and Liu Zhipeng

*Department of Physics, School of Science, Tianjin University, Tianjin 300072, China*

**Abstract**

We investigate the physical property of the $\kappa$ parameter and the $\kappa$-distribution in the $\kappa$-deformed statistics, based on Kaniadakis entropy, for a relativistic gas in an electromagnetic field. We derive two relations for the relativistic gas in the framework of $\kappa$-deformed statistics, which describe the physical situation represented by the relativistic $\kappa$-distribution function, provide a reasonable connection between the parameter $\kappa$, the temperature four-gradient and the four-vector potential gradient, and thus present for the case $\kappa \neq 0$ one clearly physical meaning. It is shown that such a physical situation is a meta-equilibrium state, but has a new physical characteristic.

---

[*] Corresponding author: jiulindu@yahoo.com.cn

In recent years, some new statistics have been put forward to generalize the classical Boltzmann–Gibbs (BG) one, such as the nonextensive statistics based on Tsallis entropy ($q$-entropy)[1], the so-called $\kappa$-deformed statistics based on the $\kappa$-entropy given by Kaniadakis [2] besides others. We mention that the concept of $\kappa$-distribution was introduced about four decades ago and also discussed recently by Leubner [3], which was actually equivalent to the nonextensive $q$-distribution, but different from Kaniadakis one. These statistics are studied all dependent on one parameter. For instance, the nonextensive statistics depends on the parameter $q$ different from unity and it will recover BG one if we take the parameter to be unity; the $\kappa$-deformed statistics depends on the parameter $\kappa$ different from zero and it will recover BG one if we take the parameter to be zero. Then it is naturally important question for us to ask what should these parameters stand for and under what physical situations should these statistics be suitable for the statistical description of a system.

For the nonextensive statistics, we know that the parameter $q$ different from unit is related to the temperature gradient, $\nabla T$, and the long-range potentials, $\varphi$, of the systems such as self-gravitating system [4,5] and plasma system [6], by $\nabla T \sim (1-q)\nabla \varphi$. Thus it can be reasonably applied to describe thermodynamic properties of the systems under an external field when they are in the nonequilibrium stationary-state. And this characteristic for the nonextensive statistics has recently received the experiment support by the helioseismological measurements [7]. As potential astrophysical examples with regard to the above theoretical investigation, a new application of the nonextensive theory for both dark matter [8], stars [9] and hot plasma [3, 10] density distributions in clustered astrophysical structures was investigated. In this situation, the potential gradient is parallel to the temperature gradient, thus fitting perfectly well into the discussions [4-7].

For the $\kappa$-deformed statistics, we actually know little about the physical property of the parameter $\kappa$ different from zero and the corresponding statistics although it has been studied by Kaniadakis and his co-authors in many papers. In this letter, we study the physical properties in the case of parameter $\kappa \neq 0$ for a relativistic gas under an electromagnetic field.

The general form of the $\kappa$-entropy proposed by Kaniadakis can be written by the $\kappa$-distribution function, $f$, [2] as

$$K(f) = a_1 S_\kappa(a_2 f) + a_3, \tag{1}$$

being $S_\kappa(f) = -\int d^3v\, f \ln_\kappa f$, which can have different expressions after having given specific choices for the parameters $a_i$ with $i=1,2,3$, where $a_i$ may depend on the position $x$. For instance, a particular case of $K(f)$ is

$$S_\kappa = -\int d^3v \left( \frac{z^\kappa}{2\kappa(1+\kappa)} f^{1+\kappa} - \frac{z^{-\kappa}}{2\kappa(1-\kappa)} f^{1-\kappa} \right), \tag{2}$$

which recovers to the Boltzmann-Gibbs entropy in the limit $\kappa \to 0$, where z is a positive real parameter that needs to be specified. Based on Kaniadakis entropy, the $\kappa$-deformed statistics can be defined by the so-called $\kappa$-exponential and $\kappa$-logarithm functions as

$$\exp_\kappa(f) = \left( \sqrt{1+\kappa^2 f^2} + \kappa f \right)^{1/\kappa}, \tag{3}$$

$$\ln_\kappa(f) = \frac{f^\kappa - f^{-\kappa}}{2\kappa}. \tag{4}$$

As one may check, we have $\exp_\kappa(\ln_\kappa(f)) = \ln_\kappa(\exp_\kappa(f)) = f$ and the above functions could reduce to the standard exponential and logarithm ones if we let $\kappa \to 0$. Actually, this $\kappa$

-framework can lead to a class of one parameter deformed structures with interesting mathematical properties[11].

We consider a relativistic gas containing $N$ point particles of mass $m$ enclosed in a volume $V$, under the action of an external four-force field $F^\mu$. The temporal evolution of the relativistic distribution function $f(x, p)$ is governed by the following relativistic $\kappa$-transport equation [11,12]:

$$p^\mu \partial_\mu f + mF^\mu \frac{\partial f}{\partial p^\mu} = C_\kappa(f), \qquad (5)$$

where $\mu = 0, 1, 2, 3$, the particles have four-momentum $p \equiv p^\mu = (E/c, \mathbf{p})$ in each point $x \equiv x^\mu = (ct, \mathbf{r})$ of the space-time, with their energy $E/c = \sqrt{\mathbf{p}^2 + m^2c^2}$, and $\partial_\mu = (c^{-1}\partial_t, \nabla)$ denotes the differentiation with respect to time-space coordinates and $C_\kappa$ the relativistic $\kappa$-collisional term. According to the generalized $H$ theorem [11,12], the relativistic version of $\kappa$-distribution function in this framework approach to

$$f(x,p) = B\exp_\kappa \theta = B\left(\sqrt{1+\kappa^2\theta^2} + \kappa\theta\right)^{1/\kappa}, \qquad (6)$$

where B may depend on $x$, and

$$\theta = \alpha(x) + \beta_\mu(x) p^\mu \qquad (7)$$

with $\alpha(x)$ a scalar and $\beta_\mu$ a four-vector, arbitrary space and time-dependent parameters. And the relativistic $\kappa$-collisional term vanishes and Eq.(5) becomes

$$p^\mu \partial_\mu f + mF^\mu \frac{\partial f}{\partial p^\mu} = 0. \qquad (8)$$

The distribution (6) can also be obtained by maximizing $K(f)$ under the right constraints. Let us now consider the external four-force field as the Lorentz one, namely,

$$F(x,p) = -(Q/mc)F^{\mu\nu}(x)p_\nu \tag{9}$$

with $Q$ the charge of the particle and $F^{\mu\nu}$ the Maxwell electromagnetic tensor. In this case, Eq.(8) is

$$p^\mu \partial_\mu f - Qc^{-1}F^{\mu\nu}p_\nu \frac{\partial f}{\partial p^\mu} = 0 \tag{10}$$

or, equivalently, it can be written as

$$p^\mu \partial_\mu f^\kappa - Qc^{-1}F^{\mu\nu}p_\nu \frac{\partial f^\kappa}{\partial p^\mu} = 0. \tag{11}$$

And the $\kappa$-distribution function, Eq.(6), [11, 12] reads

$$f(x,p) = B\exp_\kappa\left[\frac{u - [p^\mu + c^{-1}QA^\mu(x)]U_\mu}{k_B T}\right]$$

$$= B\left\{\sqrt{1+\kappa^2\left[\frac{u - [p^\mu + c^{-1}QA^\mu(x)]U_\mu}{k_B T}\right]^2} + \kappa\left[\frac{u - [p^\mu + c^{-1}QA^\mu(x)]U_\mu}{k_B T}\right]\right\}^{1/\kappa}, \tag{12}$$

where $k_B$ is Boltzmann constant, $U_\mu$ is the mean four-vector velocity of the gas with $U_\mu U^\mu = c^2$, $T(x)$ is the temperature field, $u$ is the Gibbs function per particle, and $A^\mu(x)$ is the four-vector potential. In the limit $\kappa \to 0$, it reduces to the well known relativistic expression [13] correctly,

$$f(x,p) = B\exp\left[\frac{u - [p^\mu + c^{-1}QA^\mu(x)]U_\mu}{k_B T}\right] \tag{13}$$

From Eq.(12), we have

$$f^\kappa(x,p) = B^\kappa\left\{\sqrt{1+\kappa^2\left[\frac{u - [p^\mu + c^{-1}QA^\mu(x)]U_\mu}{k_B T}\right]^2} + \kappa\left[\frac{u - [p^\mu + c^{-1}QA^\mu(x)]U_\mu}{k_B T}\right]\right\}, \tag{14}$$

and

$$f^{2\kappa} = B^{2\kappa} + 2B^{\kappa}\kappa \frac{u - (p^{\mu} + c^{-1}QA^{\mu})U_{\mu}}{k_B T} f^{\kappa} \quad . \tag{15}$$

Substitute Eq.(15) into Eq(11) and after a series of calculations, we get the equation

$$\begin{aligned}
&\left(p^{\mu}\partial_{\mu}B^{\kappa}\right)^2 + \left(p^{\mu}\partial_{\mu}B^{\kappa}\right)^2 \kappa^2 \frac{u^2}{(k_B T)^2} - 2\left(p^{\mu}\partial_{\mu}B^{\kappa}\right)^2 \frac{\kappa^2}{(k_B T)^2} u p^{\mu}U_{\mu} - 2\left(p^{\mu}\partial_{\mu}B^{\kappa}\right)^2 \frac{\kappa^2}{(k_B T)^2} u c^{-1}QA^{\mu}U_{\mu} \\
&+ \left(p^{\mu}\partial_{\mu}B^{\kappa}\right)^2 \frac{\kappa^2}{(k_B T)^2} \left(p^{\mu}U_{\mu}\right)^2 + 2\left(p^{\mu}\partial_{\mu}B^{\kappa}\right)^2 \frac{\kappa^2}{(k_B T)^2} p^{\mu}U_{\mu} c^{-1}QA^{\mu}U_{\mu} + \left(p^{\mu}\partial_{\mu}B^{\kappa}\right)^2 \frac{\kappa^2}{(k_B T)^2} \left(c^{-1}QA^{\mu}U_{\mu}\right)^2 \\
&- B^{2\kappa}\kappa^2 \frac{\left(p^{\mu}\partial_{\mu}T\right)^2}{T^2} \frac{u^2}{(k_B T)^2} + 2B^{2\kappa}\kappa^2 \frac{\left(p^{\mu}\partial_{\mu}T\right)^2}{T^2} \frac{u p^{\mu}U_{\mu}}{(k_B T)^2} + 2B^{2\kappa}\kappa^2 \frac{\left(p^{\mu}\partial_{\mu}T\right)^2}{T^2} \frac{u c^{-1}QA^{\mu}U_{\mu}}{(k_B T)^2} \\
&- B^{2\kappa}\kappa^2 \frac{\left(p^{\mu}\partial_{\mu}T\right)^2}{T^2} \frac{\left(p^{\mu}U_{\mu}\right)^2}{(k_B T)^2} - 2B^{2\kappa}\kappa^2 \frac{\left(p^{\mu}\partial_{\mu}T\right)^2}{T^2} \frac{p^{\mu}U_{\mu} c^{-1}QA^{\mu}U_{\mu}}{(k_B T)^2} - B^{2\kappa}\kappa^2 \frac{\left(p^{\mu}\partial_{\mu}T\right)^2}{T^2} \frac{\left(c^{-1}QA^{\mu}U_{\mu}\right)^2}{(k_B T)^2} \\
&- B^{2\kappa}\kappa^2 \left(\frac{c^{-1}Qp^{\mu}\partial_{\mu}A^{\mu}U_{\mu}}{k_B T}\right)^2 - B^{2\kappa}\kappa^2 \frac{Q^2}{c^2}\left(F^{\mu\nu}p_{\nu}\frac{U_{\mu}}{k_B T}\right)^2 - 2B^{2\kappa}\kappa^2 \frac{p^{\mu}\partial_{\mu}T}{T} \frac{u}{k_B T} \frac{c^{-1}Qp^{\mu}\partial_{\mu}A^{\mu}U_{\mu}}{k_B T} \\
&+ 2B^{2\kappa}\kappa^2 \frac{p^{\mu}\partial_{\mu}T}{T} \frac{p^{\mu}U_{\mu}}{k_B T} \frac{c^{-1}Qp^{\mu}\partial_{\mu}A^{\mu}U_{\mu}}{k_B T} + 2B^{2\kappa}\kappa^2 \frac{p^{\mu}\partial_{\mu}T}{T} \frac{c^{-1}QA^{\mu}U_{\mu}}{k_B T} \frac{c^{-1}Qp^{\mu}\partial_{\mu}A^{\mu}U_{\mu}}{k_B T} \\
&+ 2B^{2\kappa}\kappa^2 \frac{p^{\mu}\partial_{\mu}T}{T} \frac{u}{k_B T} \frac{Q}{c} F^{\mu\nu}p_{\nu}\frac{U_{\mu}}{k_B T} - 2B^{2\kappa}\kappa^2 \frac{p^{\mu}\partial_{\mu}T}{T} \frac{p^{\mu}U_{\mu}}{k_B T} \frac{Q}{c} F^{\mu\nu}p_{\nu}\frac{U_{\mu}}{k_B T} \\
&- 2B^{2\kappa}\kappa^2 \frac{p^{\mu}\partial_{\mu}T}{T} \frac{c^{-1}QA^{\mu}U_{\mu}}{k_B T} \frac{Q}{c} F^{\mu\nu}p_{\nu}\frac{U_{\mu}}{k_B T} + 2B^{2\kappa}\kappa^2 \frac{c^{-1}Qp^{\mu}\partial_{\mu}A^{\mu}U_{\mu}}{k_B T} \frac{Q}{c} F^{\mu\nu}p_{\nu}\frac{U_{\mu}}{k_B T} = 0.
\end{aligned} \tag{16}$$

We consider that Eq. (16) is the identically null for any arbitrary $p$ and the sum of the coefficients of every power for $p$ in this equation must be zero. Thus, when we consider the coefficients for the forth-power terms of $p$ in Eq. (16), we obtain the relation,

$$\left(\frac{\partial_{\mu}B^{\kappa}}{B^{\kappa}}\right)^2 = \left(\frac{\partial_{\mu}T}{T}\right)^2 \quad . \tag{17}$$

The coefficients of the third power of $p$ is

$$-2\left(\partial_\mu B^\kappa\right)^2 \frac{\kappa^2}{\left(k_B T\right)^2} u U_\mu + 2\left(\partial_\mu B^\kappa\right)^2 \frac{\kappa^2}{\left(k_B T\right)^2} U_\mu c^{-1} Q A^\mu U^\mu + 2 B^{2\kappa} \kappa^2 \frac{\left(\partial_\mu T\right)^2}{T^2} \frac{u U_\mu}{\left(k_B T\right)^2}$$

$$-2 B^{2\kappa} \kappa^2 \frac{\left(\partial_\mu T\right)^2}{T^2} \frac{U_\mu c^{-1} Q A^\mu U_\mu}{\left(k_B T\right)^2} + 2 B^{2\kappa} \kappa^2 \frac{\partial_\mu T}{T} \frac{U_\mu}{k_B T} \frac{c^{-1} Q \partial_\mu A^\mu U_\mu}{k_B T}$$

$$+2 B^{2\kappa} \kappa^2 \frac{\partial_\mu T}{T} \frac{U_\mu}{k_B T} \frac{Q}{c} F^{\mu\nu} \frac{U_\mu}{k_B T} = 0. \tag{18}$$

Substitute Eq.(17) into Eq.(18), we get

$$\partial_\mu T \left(\partial_\mu A^\nu - F^{\mu\nu}\right) U_\mu = 0 \tag{19}$$

The coefficients of the second power of $p$ read

$$\left(\partial_\mu B^\kappa\right)^2 + \left(\partial_\mu B^\kappa\right)^2 \kappa^2 \frac{u^2}{\left(k_B T\right)^2} - 2\left(\partial_\mu B^\kappa\right) \frac{\kappa^2}{\left(k_B T\right)^2} u c^{-1} Q A^\mu U_\mu + \left(\partial_\mu B^\kappa\right)^2 \frac{\kappa^2}{\left(k_B T\right)^2} \left(c^{-1} Q A^\mu U_\mu\right)^2$$

$$-B^{2\kappa} \kappa^2 \frac{\left(\partial_\mu T\right)^2}{T^2} \frac{u^2}{\left(k_B T\right)^2} + 2 B^{2\kappa} \kappa^2 \frac{\left(\partial_\mu T\right)^2}{T^2} \frac{u c^{-1} Q A^\mu U_\mu}{\left(k_B T\right)^2} - B^{2\kappa} \kappa^2 \frac{\left(\partial_\mu T\right)^2}{T^2} \frac{\left(c^{-1} Q A^\mu U_\mu\right)^2}{\left(k_B T\right)^2}$$

$$-B^{2\kappa} \kappa^2 \left(\frac{c^{-1} Q \partial_\mu A^\mu U_\mu}{k_B T}\right)^2 - B^{2\kappa} \kappa^2 \frac{Q^2}{c^2} \left(F^{\mu\nu} \frac{U_\mu}{k_B T}\right)^2 - 2 B^{2\kappa} \kappa^2 \frac{\partial_\mu T}{T} \frac{u}{k_B T} \frac{c^{-1} Q \partial_\mu A^\mu U_\mu}{k_B T}$$

$$+2 B^{2\kappa} \kappa^2 \frac{\partial_\mu T}{T} \frac{c^{-1} Q A^\mu U_\mu}{k_B T} \frac{c^{-1} Q \partial_\mu A^\mu U_\mu}{k_B T} + 2 B^{2\kappa} \kappa^2 \frac{\partial_\mu T}{T} \frac{u}{k_B T} \frac{Q}{c} F^{\mu\nu} \frac{U_\mu}{k_B T}$$

$$-2 B^{2\kappa} \kappa^2 \frac{\partial_\mu T}{T} \frac{c^{-1} Q A^\mu U_\mu}{k_B T} \frac{Q}{c} F^{\mu\nu} \frac{U_\mu}{k_B T} + 2 B^{2\kappa} \kappa^2 \frac{c^{-1} Q \partial_\mu A^\mu U_\mu}{k_B T} \frac{Q}{c} F^{\mu\nu} \frac{U_\mu}{k_B T} = 0. \tag{20}$$

Substitute Eqs.(17) and (19) into Eq.(20), we find

$$k_B^2 \left(\partial_\mu T\right)^2 = \kappa^2 \frac{Q^2}{c^2} \left[\left(\partial_\mu A^\nu - F^{\mu\nu}\right) U_\mu\right]^2, \tag{21}$$

which easily leads to the relation

$$k_B \left|\partial_\mu T\right| = \kappa Q \left|\partial_\mu A^\nu - F^{\mu\nu}\right|. \tag{22}$$

Furthermore, using the electromagnetic tensor expression, $F^{\mu\nu} = \partial_\mu A^\nu - \partial_\nu A^\mu$, we can write Eqs.(19) and (22), respectively, as

$$(\partial_\mu T)(\partial_\nu A^\mu) = 0 \tag{23}$$

and

$$k_B |\partial_\mu T| = \kappa Q |\partial_\nu A^\mu|, \text{ or } \kappa = k_B |\partial_\mu T| / Q |\partial_\nu A^\mu|. \tag{24}$$

where $\partial_\mu T$ and $\partial_\nu A^\mu$ are the relativistic expressions of the temperature gradient and the external potential gradient, respectively. From Eq.(24) we find that the parameter $\kappa$ is different from zero if and only if the quantity $\partial_\mu T$ is not equal to zero, thus showing that $\kappa \neq 0$ represents the physical situation of the relativistic gas with a time-spatial inhomogeneous temperature field and under the electromagnetic field. This is a nonequilibrium stationary-state.

But such a nonequilibrium relativistic gas should have one additional characteristic: from Eq.(23) we find that the temperature four-gradient $\partial_\mu T$ must be vertical to the four-potential gradient $\partial_\nu A^\mu$.

If take $\kappa = 0$, one has $\partial_\mu T = 0$, which just is the physical situation represented by the standard distribution (13). The non-relativistic versions can be written directly from Eqs.(23) and (24) by

$$\nabla T \cdot \nabla \varphi = 0; \quad k_B |\nabla T| = \kappa Q |\nabla \varphi|, \tag{25}$$

which are in agreement on the relations derived recently for the parameter $\kappa \neq 0$ in a nonequilibrium system [14]. Thus Eqs.(23) and (24) extend these classical relations to the special relativistic case.

In summary, we obtain two relations, Eqs.(23) and (24), of the parameter $\kappa$ in the $\kappa$-deformed statistics for a relativistic gas in an electromagnetic field. They describe the physical situation represented by the relativistic $\kappa$-distribution function (12) and provide a reasonable connection between the parameter $\kappa$, the temperature four-gradient and the four-vector potential gradient, so presenting the case of the parameter $\kappa \neq 0$ one clearly physical meaning. Such a physical situation is a nonequilibrium stationary-state, but has a new characteristic given by Eq.(23).

Additionally, from Eq.(17) it follows that when B is a constant also the temperature turn out to be constant, which is independent on the parameter $\kappa$. Therefore the equilibrium state emerges also in the case of a relativistic gas described by the $\kappa$-distribution based on the entropy $K(f)$ in presence of an external field. Conversely when B depends on $x$, the system admits the above nonequilibrium stationary-state. From the view of this point, the physical situation described by the $\kappa$-distribution may be called a metaequilibrium state.

*Additional remarks*: Most recently, Plastino et al reported a general discussion on the generalized entropy and the maximum entropy approach [15].

**Acknowledgements**

This work is supported by the project of "985" program of TJU of China and also by the National Natural Science Foundation of China, No. 10675088.